\documentclass[reprint,amsmath,amssymb,aps,pra]{revtex4-2}

\usepackage[utf8]{inputenc}
\usepackage{graphicx}
\usepackage{float}
\usepackage{dcolumn}
\usepackage{caption}
\usepackage{bm}
\usepackage{amsmath}

\usepackage{threeparttable}
\usepackage{tabularx, booktabs}
\usepackage{multirow}

\usepackage{color}
\usepackage{upgreek}

\newcolumntype{Y}{>{\centering\arraybackslash}X}

\newcolumntype{L}[1]{>{\raggedright\let\newline\\\arraybackslash\hspace{0pt}}m{#1}}
\newcolumntype{C}[1]{>{\centering\let\newline\\\arraybackslash\hspace{0pt}}m{#1}}
\newcolumntype{R}[1]{>{\raggedleft\let\newline\\\arraybackslash\hspace{0pt}}m{#1}}

\begin{document}

\title{Pauli principle in polaritonic chemistry}

\author{Tam\'as Szidarovszky}
    \email{tamas.janos.szidarovszky@ttk.elte.hu}
    \affiliation{%
    Institute of Chemistry,
	ELTE E\"otv\"os Lor\'and University, H-1117 Budapest, P\'azm\'any P\'eter s\'et\'any 1/A, Hungary}%


\begin{abstract}
The consequences of enforcing permutational symmetry, as required by the Pauli principle (spin-statistical theorem), on the state space of molecular ensembles interacting with the quantized radiation mode of a cavity are discussed.
The Pauli-allowed collective states are obtained by means of group theory, i.e., by projecting the state space onto the appropriate irreducible representations of the permutation group of the indistinguishable molecules.
It is shown that with increasing number of molecules the ratio of Pauli-allowed collective states decreases very rapidly.
Bosonic states are more abundant than fermionic states, and the brightness of Pauli-allowed state space (the contribution from photon excited states) increases(decreases) with increasing fine structure in the energy levels of the material ground(excited) state manifold.
Numerical results are shown for the realistic example of rovibrating H$_2$O molecules interacting with an infrared cavity mode.
\end{abstract}

\maketitle
% In Ref. \cite{23PeKoStZh} permutational symmetries were exploited to drastically reduce the computational cost of ab initio quantum dynamics simulations for many molecules interacting with a cavity mode, and to introduce an effective single molecule model to approximately capture the dynamics of the entire ensemble.

\section{Introduction}

The Pauli principle, also called the spin-statistical theorem, is a fundamental restriction on the permutational symmetry of the wave functions of quantum systems \cite{39Fi,40Pa,06BuJe}, having a huge impact on the physicochemical properties of matter.
In a chemical context, antisymmetrization of the wave function with respect to electron permutations, as required by the Pauli principle, turns a Hartree product into a Slater determinant.
%directly giving rise to electron exchange energy \cite{06Jensen}.
The Pauli principle also restricts the space of physically allowed quantum states, i.e., from all the possible eigenstates of the system Hamiltonian only those are realized in nature which satisfy the required permutational symmetry.
This is the reason, for example, why the lowest-energy state of the Li atom (three-electron Hamiltonian) is not realized physically (it is Pauli forbidden) \cite{95KlLuBa}.
In a similar fashion, applying the Pauli principle to the identical atomic nuclei in molecules (i) causes the lowest-energy rovibrational eigenstate of H$_3^+$ to be Pauli forbidden \cite{13FuSzMaFa}, and (ii) gives rise to nuclear spin-statistical weights \cite{06BuJe}, which fundamentally contribute to the structure of molecular infrared (IR) and microwave spectra, as well as thermochemistry.
In this work we investigate the direct effects of the Pauli principle on polaritonic chemistry, which studies the properties and dynamics of molecules interacting with quantized radiation modes \cite{Ebbesen_review_2016,Zhou_review_2018,Feist_GarciaVidal_ACSphotonics_2018,Edina1,Herrera_PRL_2016,Hertzog2019,Herrera_2020,Kowalewski2017}.

In polaritonic chemistry molecules and the cavity mode are usually considered to be in the (ultra)strong coupling regime, i.e., light-matter coupling is assumed to be larger than the cavity leakage. This leads to the formation of so-called polaritons: coherent superposition states having both material- and photonic-excited components \cite{Ebbesen_review_2016,Hertzog2019,Herrera_2020}.
Depending on the cavity-mode wavelength, the confined photonic modes of the cavity can efficiently couple with either electronic or (ro)vibrational molecular states, leading to electronic or (ro)vibrational polaritons, respectively.
Because a single radiation mode can simultaneously interact with multiple molecules, and the light-matter interaction can also change the state of the field, an indirect interaction is formed between the molecules, introducing so-called collective effects \cite{Ebbesen_review_2016,Herrera_2020}.
Collective effects play a central role in polaritonic chemistry, for example, they are responsible for the well-known $\sqrt{n}$ scaling of the light-matter coupling strength when $n$ molecules interact with the cavity mode.
Furthermore, when $n$ identical molecules interact with a cavity mode, collective states can be formed, which are coherent superpositions of different material- and photonic excitations.
The first excited manifold in principle contains, in addition to the two bright (upper and lower) polaritonic states, $(n-1)$ so-called dark states  \cite{Herrera_2020,Herrera2017,Vendrell2018,Xiang2018,18HeSp,22CsVeHaVi}.
Although it is debated whether quantum coherence on a mesoscopic scale can indeed be realized in IR microcavities \cite{Sidler2022}, the existence (of mesoscopic amounts) of dark states has been a key factor in considering and describing the physicochemical properties and reactions of vibropolaritonic systems \cite{Herrera2017,Vendrell2018,Xiang2018,18HeSp,CamposGonzalezAngulo_2020,Du2022}.
In the simple model of two two-level systems in a resonant cavity, the first excited manifold contains the two bright polaritonic states
$\vert \Psi _\pm \rangle \propto (\vert e \rangle \vert g \rangle \vert 0 \rangle + \vert g \rangle \vert e \rangle \vert 0 \rangle \pm \sqrt{2} \vert g \rangle \vert g \rangle \vert 1 \rangle)$, and the dark state $\vert \Psi _{\rm d} \rangle \propto \vert e \rangle \vert g \rangle \vert 0 \rangle - \vert g \rangle \vert e \rangle \vert 0 \rangle $,
where $\vert g \rangle$ and $\vert e \rangle$ are the ground and excited material states, respectively, while $\vert 0 \rangle$ and $\vert 1 \rangle$ are photon number states.

Returning to the Pauli principle, permutational symmetry with respect to the electrons is implicitly incorporated in the electronic structure methods of polaritonic chemistry \cite{Tokatly2013,Ruggenthaler2014,Malave2022,Schfer2021,Haugland2020,Haugland2021} and the invariance of the wave function with respect to the permutation of emitters/molecules has also be exploited in some theoretical works of the field \cite{22Ce, 18ZeKiKe, 18HeSp, 20Sp, 23PeKoStZh}, primarily to reduce computational cost by reducing the basis set size needed to describe the polaritons and to arrive to effective single-molecule models.
It also has been shown that, even if molecular indistinguishability is not considered, permutational symmetry with respect to the exchange of molecules can play a significant role in the physicochemical properties of systems forming rovibrational polaritons \cite{21SzBaHaVi}.
In addition, molecules can be both fermions or bosons, depending on their total (nuclear and electronic) spin.
Therefore, the symmetry of collective states with respect to the permutation of the indistinguishable molecules can be both symmetric or antisymmetric.
In the example given above, only the $\vert \Psi _\pm \rangle$ bright polaritons can exist for bosons and only the $\vert \Psi _{\rm d} \rangle$ dark polariton can exist for fermions.
This raises the central questions of this paper: \textit{(i)What are the differences in the physically allowed state space of polariton formation for bosonic and fermionic molecules?
(ii) To what extent can bright- and dark polaritons coexist?
(iii) Is it realistic to assume that there is a mesoscopic amount of Pauli-allowed collective states in a cavity setting?}

\section{Theory}

It is important to discuss, at this point, the assumptions and limitations of the theory to be formulated below.
The questions raised at the end of the previous section are addressed assuming that (a) it is appropriate to consider the permutational symmetry only with respect to the full molecules and that (b) the molecules are indeed indistinguishable.

Naturally, if the permutational symmetry of all the electrons and nuclei are considered, this automatically leads to the correct permutational symmetry with respect to exchanging full molecules.
The line of reasoning to consider only permutation with respect to full molecules is similar to that used in theoretical molecular spectroscopy, when one uses the molecular symmetry (MS) group instead of the complete nuclear permutation and inversion (CNPI) group \cite{06BuJe}.
The MS group is a subgroup of the CNPI group, which contiains those symmetry operations, which are physically feasible for the system under investigation. For example, when carrying out a simulation on a gas sample of strongly-bound molecules at room temperature, symmetry operations in the CNPI group that would involve bond breaking are omitted from the MS group.
In a polaritonic chemistry setting, intermolecular exchange of individual electrons or nuclei is assumed to be unlikely for most molecules, therefore, these symmetry operations are not included in the permutational symmetry group used in this work.
However, the permutation of full molecules can be physically realized, considering that the interaction with the cavity radiation can exchange the state of internal degrees of freedom (through photon emission and absorbtion), while molecules can swap places, exchanging the translational part of their wave functions.
On the other hand, for indistinguishability to play a role, (a) the translation part of the molecular wave functions should overlap, meaning that the de Broglie wavelength of the molecules should be comparable to the average molecular distance, and (b) decoherence should be small.
%The former is enhanced by reducing temperature and/or increasing the molecule density (say a liquid phase experiment where the cavity radiation is strongly coupled to the solvent, i.e., the average molecular distance is very small), however, increasing molecule density also increases intermolecular interactions and decoherence.
%I can not judge whether future experimental setups could be realized in a way that the Pauli principle indeed plays a role.

For a polaritonic system composed of the radiation modes and $n$ indistinguishable molecules, the collective polaritonic eigenstates need to transform as the [1$^n$]([$n^1$]) one-dimensional irreducible representations (irrep) of the $S_n$ symmetric group of degree $n$, whose elements permute the $n$ equivalent fermionic(bosonic) molecules \cite{95Pauncz,06BuJe}.
For [$n^1$] all characters are equal to one, while for [1$^n$] the characters are one and minus one for even and odd permutations, respectively.

Taking a single cavity mode and $n$ molecules, the state space is spanned by the $\{\vert N \rangle \vert k_{\rm 1} \rangle ... \vert k_{n} \rangle\}$ set of functions, where $N$ is the photon number and the $i$th molecule is in the state $k_{i}$. Projecting this space onto the appropriate irreducible representations of $S_n$ reveals the physically allowed space for the formation of collective polaritonic states.
The projection is carried out with standard tools of group theory \cite{06BuJe}, i.e., the projectors
\begin{equation}
 \hat{P}_{\rm fermion/boson}=\frac{1}{h}\sum_{\hat{R}}\chi ^{[1^n]/[n^1]}[\hat{R}]\hat{R}
 \label{eq_projector}
\end{equation}
are used, where $h$ is the order of the $S_n$ group, $\hat{R}$ goes over all symmetry operations (permutations) in $S_n$, and $\chi ^{\Gamma} [\hat{R}]$ is the character of $\hat{R}$ in $\Gamma$ irrep. All permutations $\hat{R}$ can be written as a product of transpositions $(ij)$, and the effect of $(ij)$ on a basis function is given by
%\begin{equation}
%\begin{aligned}
 $(ij) \vert N \rangle \vert k_{\rm 1} \rangle ... \vert k_{i} \rangle ... \vert k_{j} \rangle ... \vert k_{n} \rangle = \vert N \rangle \vert k_{\rm 1} \rangle ... \vert k_{j} \rangle ... \vert k_{i} \rangle ... \vert k_{n} \rangle$.
%\end{aligned}
%\end{equation}
For realistic molecular models within a microcavity, the state labels $k_i$ should incorporate all accessible degrees of freedom.
The group theoretical procedure above  makes no assumptions about the specific form of the system Hamiltonian,
only the state space is manipulated, which is expressed in a general way with direct product basis functions.
Therefore, the procedure can be used with molecular models of arbitrary complexity, as long as the wave function formalism is appropriate.
Extending the framework to multiple radiaton modes is also possible.
More details and a simple example about the approach outlined above can be found in the Appendix.

\section{Results and discussion}

%\subsection{First-excited manifold --}

\textit{First-excited manifold --} We start with the first-excited manifold of $n$ two-level systems interacting with a lossless cavity mode.
The state space is then spanned by the $(n+1)$ basis functions $\vert 1 \rangle \vert g \rangle ... \vert g \rangle $ and $\{\vert 0 \rangle \vert g \rangle  ... \vert e \rangle ... \vert g \rangle  \}$, where the first ket vector in the direct products is the photon number state, $g$ and $e$ stand for the material ground and excited states, respectively, and there are $n$ zero-photon states.
Table \ref{tab:paulistates} shows the number of Pauli-allowed linear combinations obtained from this set of basis functions by projection onto [$n^1$] or [1$^n$].
As shown by the numerical examples in Table \ref{tab:paulistates}, two bosonic states exist for all $n$, while there are no fermionic states for $n>2$.
The trace of the photon-number matrix $\mathbf{N}_{\rm ph}$ (the matrix representation of $\hat{a}^{\dagger}\hat{a}$, where $\hat{a}^{\dagger}$ and $\hat{a}$ are the cavity photon creation and annihilation operators, respectively) computed within a given subspace, shown in Table \ref{tab:paulistates},
reveals the number of basis functions in which the cavity mode is excited. Thus we identify ${\rm Tr}(\mathbf{N}_{\rm ph})$ as the number of bright basis function, i.e., the number of basis functions representing a photon in the cavity mode.
It can be seen in Table \ref{tab:paulistates} that the bosonic state space contains bright basis functions with photonic excitation (${\rm Tr}(\mathbf{N}_{\rm ph}>0)$), while the fermionic basis functions are dark (${\rm Tr}(\mathbf{N}_{\rm ph}=0)$), as can also be verified for $n=2$ in the simple example above.
This means that in principle no polaritons can be formed by fermionic two-level systems in the first-excited manifold.

\begin{table*}[]
	\centering
	\caption{
    \textbf{Columns 3-5}: Number of states in the first excitation manifold of $n$ molecules, having $m$ levels with $m_{\rm g}$ in the material ground state manifold, interacting with a cavity mode.
	Bosonic and fermionic subspaces were obtained by projecting the full state space onto the appropriate irreducible representations of the $S_n$ group, see text for details.
	Percentage values in parentheses show the relative number of basis functions with respect to the unsymmetrized ``no Pauli'' case.
	\textbf{Columns 6-8}: Number of bright basis functions in the given subspaces, obtained as the trace of the photon number operator.
	Percentage values in parentheses show the relative number of bright basis functions with respect to the number of all the states in the given subspace.
	}\label{tab:paulistates}
	\begin{tabularx}{0.99\textwidth}{c *{8}{Y}}
	\toprule[0.2em]
%		\multirow{2}{c}{$m_{\rm g}$} & \multirow{2}{c}{$n$} &  \multicolumn{3}{c}{number of states} \\
		 & &  \multicolumn{3}{c}{number of states} & \multicolumn{3}{c}{Tr($\mathbf{N}_{\rm ph}$) = number of bright basis functions} \\
         $m_{\rm g}$ & $n$ & no Pauli  & boson & fermion & no Pauli  & boson & fermion \\
                    \midrule[0.2em]
	\multicolumn{8}{c}{2-level system} \\
                    \midrule
	1	&2	&3	&2	(67\%)	&1	(33\%)	&1	&	1	(50\%)	&	0	(0\%) \\
        &3	&4	&2	(50\%)	&0	(0\%)	&1	&	1	(50\%)	&	0	(0\%) \\
        &4	&5	&2	(40\%)	&0	(0\%)	&1	&	1	(50\%)	&	0	(0\%) \\

                    \midrule[0.2em]
	\multicolumn{8}{c}{5-level system} \\
                    \midrule
    1	& 2	& 9 	& 5    (56\%)	&4	(44\%) &1	&	1	(20\%)	&	0	(0\%) \\
        & 3	& 13	& 5    (38\%)	&0	(0\%)  &1	&	1	(20\%)	&	0	(0\%) \\
        & 4	& 17	& 5    (29\%)	&0	(0\%)  &1	&	1	(20\%)	&	0	(0\%) \\
                    \midrule
    2	& 2	& 16	& 9    (56\%)	&7	(44\%) &4	&	3	(33\%)	&	1	(14\%) \\
        & 3	& 44	& 13 	(30\%)	&3	(7\%) &8    &	4	(31\%) &	0	(0\%) \\
        & 4	& 112   & 17	(15\%)	&0	(0\%) &16	&	5	(29\%)	&	0	(0\%) \\
                    \midrule
    3	& 2	& 21	& 12 	(57\%)	&9	(43\%) &9	&	6	(50\%)	&	3	(33\%) \\
        & 3	& 81	& 22 	(27\%)	&7	(9\%) &27	&	10	(45\%)	&	1	(14\%) \\
        & 4	& 297   & 35	(12\%)	&2	(1\%) &	81	&	15	(43\%)	&	0	(0\%) \\
                    \midrule
    4	& 2	& 24	& 14 	(58\%)	&10	(42\%)&	16	&	10	(71\%)	&	6	(60\%) \\
        & 3	& 112   & 30	(27\%)	&10	(9\%) &	64	&	20	(67\%)	&	4	(40\%) \\
        & 4	& 512   & 55	(11\%)	&5	(1\%) &	256	&	35	(64\%)	&	1	(20\%) \\
                    \midrule[0.2em]
	\multicolumn{8}{c}{10-level system} \\
                    \midrule
	1  & 2	&19	    &10	    (53\%)	&9	(47\%) &	1	&	1	(10\%)	&	0	(0\%) \\
       & 3	&28	    &10	    (36\%)	&0	(0\%) &	    1	&	1	(10\%)	&	0	(0\%) \\
       & 4	&37	    &10	    (27\%)	&0	(0\%) &	    1	&	1	(10\%)	&	0	(0\%) \\
                    \midrule
    3  & 2	& 51   &	27  (53\%)	&	24 (47\%) &	9	&	6	(22\%)	&	3	(13\%) \\
       & 3	& 216   &   52 (24\%)	&	22 (10\%) &	27	&	10	(19\%)	&	1	(5\%) \\
       & 4	& 837   &   85 (10\%)	&	7 (1\%) &	81	&	15	(18\%)	&	0	(0\%) \\
                    \midrule
    5  & 2	&75	    &40	    (53\%)	&35	(47\%) &	25	&	15	(38\%)	&	10	(29\%) \\
       & 3	&500    &110    (22\%)	&60	(12\%) &	125	&	35	(32\%)	&	10	(17\%) \\
       & 4	&3125   &245    (8\%)	&55	(2\%) &	    625	&	70	(29\%)	&	5	(9\%) \\
                    \midrule
    7  & 2	&	91    &	49    (54\%)	&42	(46\%) &	49	&	28	(57\%)	&	21	(50\%) \\
       & 3	&   784 & 168   (21\%)	&98	(13\%) &	343	&	84	(50\%)	&	35	(36\%) \\
       & 4	&   6517& 462   (7\%)	&140	(2\%) &	2401  	& 210	(45\%)	&	35	(25\%) \\
%                    \midrule
%    9  & 2	&99	    &54	    (55\%)	&45	(45\%) &	81	&	45	(83\%)	&	36	(80\%) \\
%       & 3	&972    &210    (22\%)	&120	(12\%)&	729	&	165	(79\%)	&	84	(70\%) \\
%       & 4	&9477   &660    (7\%)	&210	(2\%) &	6561&	495	(75\%)	&	126	(60\%) \\

	\bottomrule[0.2em]
	\end{tabularx}
\end{table*}

However, molecules are not two-level systems; therefore, we now turn to more complicated cases and investigate the first-excited manifold of $n$ number of $m$-level systems interacting with a lossless cavity mode.
The state space is then spanned by the basis functions $\{\vert 1 \rangle \vert k_{\rm 1} \rangle ... \vert k_{n} \rangle\}_{k_i=1}^{m_{\rm g}}$ and $\{\vert 0 \rangle \vert k_{\rm 1} \rangle ... \vert k_{n} \rangle\}_{k_i=m_{\rm g}+1}^{m}$, where
%the first ket vector is the photon number state and
the lowest $m_{\rm g}$ energies of the $m$-level systems are categorized to be in the molecular ground state manifold, while the eigenstates $m_{\rm g}+1$ to $m$ are categorized as excited states.
Examples for such a grouping of molecular levels could be the ground- and first-excited vibrational (electronic) states with their respective rotational (rovibrational) fine structure.
Table \ref{tab:paulistates} summarizes the results for the $m=5$ and $m=10$ systems and various $m_{\rm g}$ values.
The relative number of both bosonic and fermionic basis functions, with respect to the unsymmetrized basis set size, rapidly decreases with increasing molecule number.
Irrespective of the $m$ number of levels, the number of bosonic states is larger than the fermionic states, and in fact for $m_{\rm g}=1$, no fermionic state exists for $n>2$.
% For $m_{\rm g}>1$, Pauli-allowed fermionic states can be formed, and with increasing $m_{\rm g}$ the sets of both bosonic and fermionic states become brighter, as demonstrated by the photon number trace shown in Table ... .
The photon number trace shown in Table \ref{tab:paulistates} demonstrates that with increasing fine structure in the ground state manifold, i.e. with increasing $m_{\rm g}$, the sets of both bosonic and fermionic basis functions become brighter, because more bright Pauli-allowed combinations of basis functions can be generated (compare $m=5$, $m_{\rm g}=2$ with $m=10$, $m_{\rm g}=7$). On the other hand, the ratio of bright basis functions decreases with increasing fine structure in the excited state manifold (compare $m=5$, $m_{\rm g}=3$ with $m=10$, $m_{\rm g}=3$).
Note that for a specific system the brightness of the polaritonic states, i.e., the degree of mixing between the bright and the dark basis functions of the Pauli-allowed state space, depends on the specific form of the Hamiltonian.

In summary, the relative number of Pauli-allowed collective states in the first-excited manifold rapidly decreases with increasing molecule number, bosonic states are more abundant than fermionic states, and the average brightness of the Pauli-allowed
state space
increases(decreases) with increasing fine structure in the energy levels of the material ground(excited) state manifold. Extending the first-excited manifold to the complete set of direct-product basis functions gives similar conclusions, as shown below.

\begin{figure}[!ht]
\centering \includegraphics[width=0.23\textwidth]{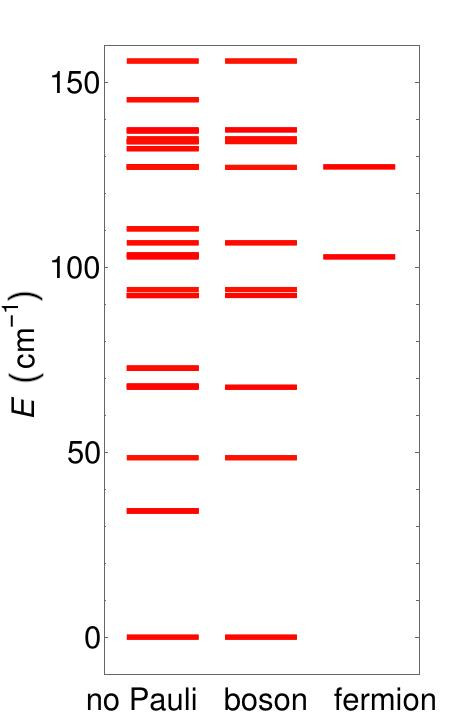}
\centering \includegraphics[width=0.24\textwidth]{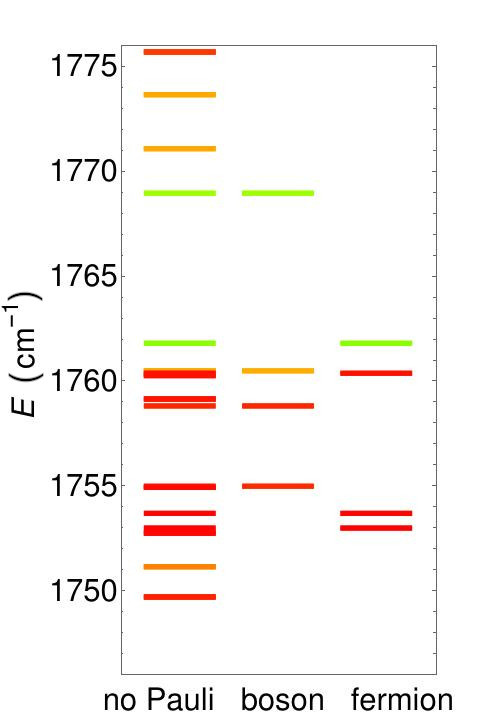}

\caption{Energy levels of the ``3$\times$H$_2$O + IR cavity mode'' model, obtained directly ('no Pauli') or after projecting the basis functions onto the [$3^1$] ('boson') or [1$^3$] ('fermion') irreducible representations of the $S_3$ permutation group, whose elements permute the indistinguishable H$_2$O molecules. The colors of the lines represent their character: red indicates zero expectation value for the photon number, while green represents one photon expectation value. Yellow indicates a mixture of photonic and material excitations, i.e., the formation of polaritonic states.}
\label{fig:en_nmol3}
\end{figure}

\textit{Pauli-allowed energetics --} The energy levels and wave functions of $n$ two-level systems interacting with a resonant cavity mode can be derived analytically \cite{Garraway2011,Ebbesen_review_2016}%, for a simple derivation regarding the first-excited manifold see the Supplementary Information
.
In agreement with Table \ref{tab:paulistates}, the first-excited manifold contains two bosonic polaritons for all $n$ and $(n-1)$ degenerate dark states, which is fermionic for $n=2$ and are Pauli forbidden for $n>2$.
Now we turn to a more realistic example of the 10-level system with $m_{\rm g}=5$, which represents the ground vibrational state and the bending fundamental of \textit{ortho}-H$_2$$^{16}$O with a fine structure of rotational levels up to $J=2$.
The accurate computation of the rovibrational polaritons of H$_2$O interacting with a near-resonant IR cavity mode was described in Ref. \cite{23Sz}.
Utilizing that approach, the polaritonic energies of three H$_2$O molecules interacting with the IR cavity mode were computed both before and after projecting the full set of basis functions onto [3$^1$] or [1$^3$].
These computations used the rigid rotor harmonic oscillator (RRHO) model of Ref. \cite{23Sz} and included the zero- and one-photon states for the cavity mode with $\tilde{\nu}=1681$ cm$^{-1}$ photon energy, nearly resonant with the RRHO $(0 1 0)[1 1 1]\leftarrow(0 0 0)[0 0 0]$ rovibrational transition, where $(n_1 n_2 n_3)[J K_a K_c]$ are the usual normal mode and asymmetric top quantum numbers \cite{06BuJe}. The light-matter coupling strength was set to $g=490$ cm$^{-1}$, which represents the coupling strength between a single photon electric field and the atomic unit of the dipole moment \cite{23Sz}.
Projecting the full set of $10\times10\times10\times2=2000$ basis functions lead to 440 bosonic and 240 fermionic basis functions.
Note that the most abundant isotopologue, H$_2$$^{16}$O, is a bosonic molecule, while the rare H$_2$$^{17}$O isotopologue is fermionic.
As can be seen in the results presented in Fig. \ref{fig:en_nmol3}, by restricting the state space to those satisfying the Pauli principle, the energy landscape changes significantly.
As can be expected from Table \ref{tab:paulistates}, the average energy level spacing increases, and in addition, for fermionic molecules the ground state energy also increases. These drastic changes in energetics should have considerable impact on the thermochemical and dynamic properties of the system.

\begin{figure}[!ht]
\centering \includegraphics[width=0.52\textwidth]{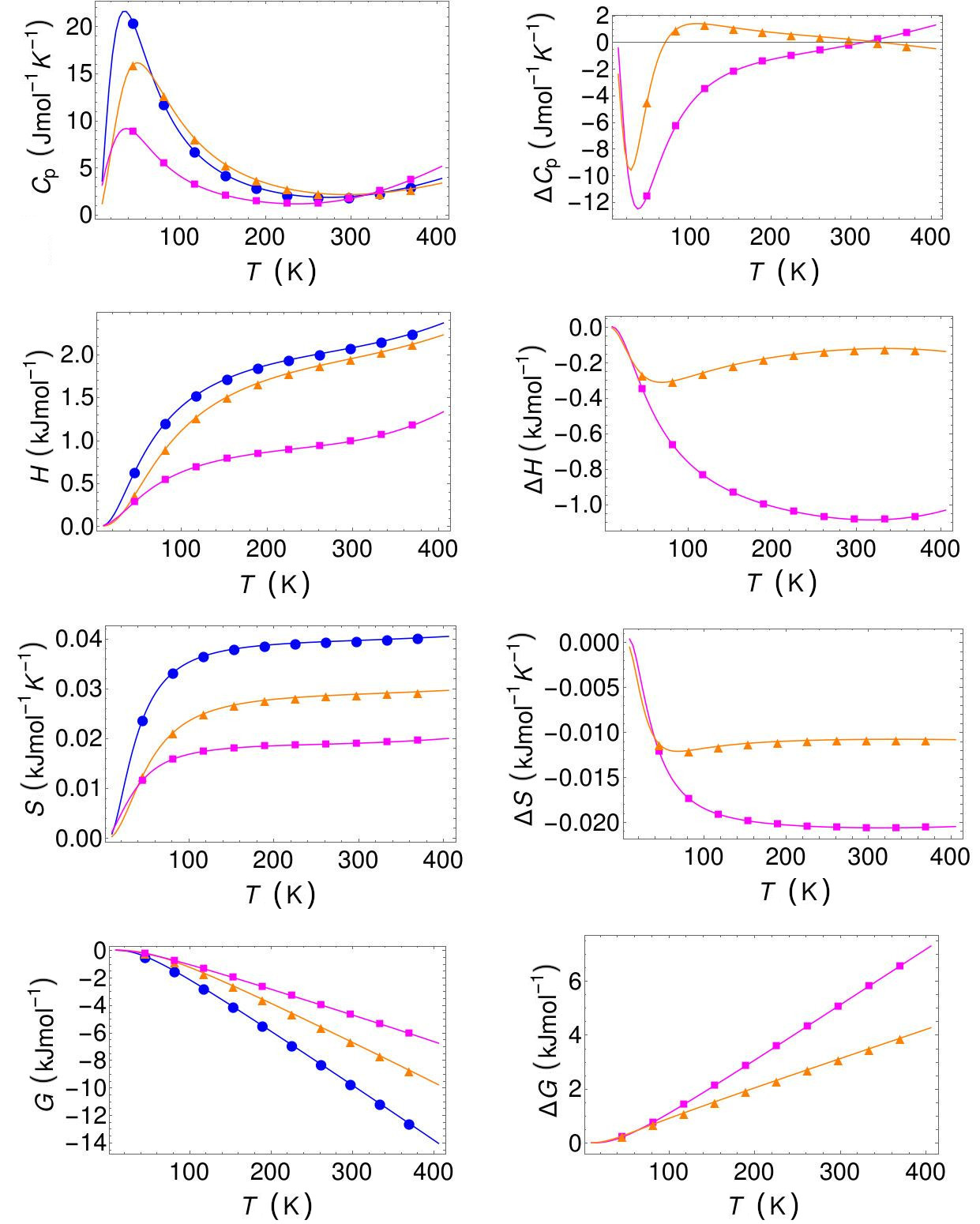}

\caption{Computed rovibrophotonic contribution to the thermodynamic properties of the ``3$\times$H$_2$O + IR cavity mode'' model, as a function of temperature $T$, obtained directly (blue circles) or after projecting the state space onto the [$3^1$] (orange triangles) or [1$^3$] (magenta squares) irreducible representations of the $S_3$ permutation group, whose elements permute the indistinguishable H$_2$O molecules. Left column: absolute values. Right column: values relative to the blue circles.}
\label{fig:thermo}
\end{figure}

With the energy levels at hand, the impact on thermochemistry can be tested by using the direct summation technique \cite{16FuSzHrKy,Pilar2020,23Fa} to compute the rovibrophotonic contribution to thermodynamic properties.
The results obtained using the formulae of Ref. \cite{16FuSzHrKy} (with the translational contributions, which are absent in our model, removed) are shown in Figure \ref{fig:thermo} and demonstrate that different temperature dependence of the thermodynamic functions is obtained from the same system Hamiltonian if different permutational symmetry is enforced on the state space.
For example, enforcing either bosonic or fermionic statistics drastically reduces the heat capacity at low temperatures, as can be expected from the reduced density of states, and increases the molar Gibbs free energy at room temperature by several kJmol$^{-1}$.

\section{Summary and concluding remarks}
It was shown that when $n$ indistinguishable molecules interact with a lossless cavity mode, the number of Pauli-allowed states relative to the full state space rapidly decreases with increasing molecule number $n$.
Based on the results of this paper, judging the magnitude of physically realized collective states in an experimental setup containing mesoscopic amount of molecules is far from trivial and should be done with caution,
given that quantum indistinguishability indeed plays a role.
The brightness (relative number of basis functions with excited cavity mode) of the Pauli-allowed
state space
increases(decreases) with increasing fine structure in the energy levels of the molecular ground(excited) state manifold.
Numerical results on the ``3$\times$H$_2$O + IR cavity mode'' model demonstrated that enforcing the permutational symmetry on the state space, as required by the Pauli principle, considerably affects the energy landscape and resulting thermodynamic properties of the system.
Because the different isotopologues of molecules can follow different spin statistics (some isotopologues might be bosons, while others fermions), the results of this work suggest that polariton formation and the resulting physicochemical properties of the system can be very different for the different isotopologues.
Therefore, designing polaritonic experiments in which quantum indistinguishability plays a role, and carrying out these experiments on different purified samples, containing only a bosonic or fermionic isotopologue, could be a useful approach to investigate issues such as the existence or role of mesoscopic collective (dark) states, and could be an additional control knob in polaritonic chemistry.

\section{Acknowledgements}
This research was supported by the János Bolyai Research Scholarship of the Hungarian Academy of Sciencesn, by the ÚNKP-22-5 New National Excellence Program of the Ministry for Innovation and Technology, and by NKFIH (Grant No. FK134291). The author is grateful to József Sike for valuable discussions.

%\bibliography{vibrational_polaritons}

\section{Appendix}

\renewcommand{\theequation}{A\arabic{equation}}
\setcounter{equation}{0}

To obtain the Pauli-allowed subspace from the space spanned by the $\{\vert N \rangle \vert k_{\rm 1} \rangle ... \vert k_{n} \rangle\}$ set of direct product functions, the following procedure was carried out.
\begin{enumerate}
 \item Because the permutations only affect the molecular states, only the molecular state space is considered initially.
 The molecular state space is generated in the uncoupled eigenstate direct-product representation, i.e., using
 \begin{equation}
 \begin{split}
  \vert 1 \rangle \hdots \vert 1 \rangle\ \leftrightarrow \left[\begin{array}{c} 1 \\ 0 \\ \vdots \\ 0 \end{array}\right]
  \otimes &\hdots \otimes
  \left[\begin{array}{c} 1 \\ 0 \\ \vdots \\ 0 \end{array}\right]
  ,\\
  \vert 2 \rangle \hdots \vert 1 \rangle\ \leftrightarrow \left[\begin{array}{c} 0 \\ 1 \\ \vdots \\ 0 \end{array}\right]
  \otimes &\hdots \otimes
  \left[\begin{array}{c} 1 \\ 0 \\ \vdots \\ 0 \end{array}\right]
  , \\
  &\vdots
  \label{eq_directproductbasis}
  \end{split}
 \end{equation}
(a) If the full molecular state space is considered, then this results in $m^n$ vectors of dimension $m^n$, where $m$ is the number of unique molecular eigenstates.

(b) If only the first-excited manifold is considered in the ``cavity+molecules'' system, then two sets of molecular basis functions are generated.
In the first set, which is to be paired with the zero-photon radiation state, one molecule is in a state associated with the excited state manifold $k_i>m_{\rm g}$, while all other molecules are in a state within the ground state manifold $k_{j\ne i}\le m_{\rm g}$, where the lowest $m_{\rm g}$ energies of the $m$-level molecules are categorized to be in the molecular ground state manifold, while the eigenstates $m_{\rm g}+1$ to $m$ are categorized as excited states.
In the second set of molecular basis functions, which is to be paired with the one-photon radiation state, all molecules are in a state associated with the ground state manifold.
 \item Each element (permutation) in the $S_n$ symmetric group is expanded as a product of transpositions $(ij)$. The effect of each transposition on the basis functions in Eq. (\ref{eq_directproductbasis}) is obtained simply by exchanging the $i$th and $j$th term in the direct products.
 \item Using the appropriate irrep of $S_n$ ( [1$^n$] or [$n^1$] for Fermions and Bosons, respectively), the projector in Eq. (\ref{eq_projector}) is acted upon the basis functions of Eq. (\ref{eq_directproductbasis}).

(a) If the full molecular state space is considered, the projection leads to $m^n$ projected molecular vectors.
Taking the direct product of these with the states of the radiation field gives the full projected state space $V_{\rm proj}$.

(b) If only the first excited manifold is considered, then the zero-photon and one-photon radiaton states are multiplied with the appropriate set of molecular basis functions (see 1.(b) above) to give $V_{\rm proj}$.

\item The number of linearly independent vectors in the projected state space (the dimension of $V_{\rm proj}$) is determined, giving the number of Pauli-allowed basis functions in the ``cavity+molecules'' system.
\end{enumerate}

\subsection{Example}
The above procedure is demonstrated for two two-level systems interacting with one cavity mode.
The character table of $S_2$ is shown in Table \ref{tab:S2}.

\begin{table*}[!ht]
	\centering
	\caption{
	Character table of the $S_2$ symmetry group}\label{tab:S2}
	\begin{tabularx}{0.49\textwidth}{c *{3}{Y}}
	\toprule[0.2em]
	& $E$ & (12) \\
	\# &  1  &  1  \\
                    \midrule
    s[2$^1$] & 1 & 1 \\
	s[1$^2$] & 1 & -1 \\
	\bottomrule[0.2em]
	\end{tabularx}
\end{table*}

\begin{enumerate}
 \item Generating the molecular state space

 (1.a) If the full molecular state space is considered, then one obtains
 \begin{equation}
 \begin{split}
  \vert g \rangle \vert g \rangle\ \leftrightarrow \left[\begin{array}{c} 1 \\ 0 \\ \end{array}\right] \otimes \left[\begin{array}{c} 1 \\ 0 \\  \end{array}\right] = \left[\begin{array}{c} 1 \\ 0 \\ 0 \\ 0 \end{array}\right]
  , \,\, \\
  \vert g \rangle \vert e \rangle\ \leftrightarrow \left[\begin{array}{c} 1 \\ 0 \\ \end{array}\right] \otimes \left[\begin{array}{c} 0 \\ 1 \\  \end{array}\right] = \left[\begin{array}{c} 0 \\ 1 \\ 0 \\ 0 \end{array}\right]
  , \,\, \\
  \vert e \rangle \vert g \rangle\ \leftrightarrow \left[\begin{array}{c} 0 \\ 1 \\ \end{array}\right] \otimes \left[\begin{array}{c} 1 \\ 0 \\  \end{array}\right] = \left[\begin{array}{c} 0 \\ 0 \\ 1 \\ 0 \end{array}\right]
  , \,\, \\
  \vert e \rangle \vert e \rangle\ \leftrightarrow \left[\begin{array}{c} 0 \\ 1 \\ \end{array}\right] \otimes \left[\begin{array}{c} 0 \\ 1 \\  \end{array}\right] = \left[\begin{array}{c} 0 \\ 0 \\ 0 \\ 1 \end{array}\right]
  . \,\,
  \end{split}
  \label{eq_2leveldirectproductbasis}
  \end{equation}
(1.b) If only the first excited manifold of the ``cavity radiation + molecules'' system is considered, then the set of molecular states associated with the ground state manifold is $\{\vert g \rangle \vert g \rangle \}$, while the set associated with the excited state manifold is $\{\vert g \rangle \vert e \rangle, \vert e \rangle \vert g \rangle \}$.

\item Effect of the permutation operators

The effect of the identity operator $E$ is trivial, and (12) gives
 \begin{equation}
 \begin{split}
  (12) \vert g \rangle \vert g \rangle\ = \vert g \rangle \vert g \rangle\  \leftrightarrow \left[\begin{array}{c} 1 \\ 0 \\ 0 \\ 0 \end{array}\right]
  , \,\, \\
  (12) \vert g \rangle \vert e \rangle\ = \vert e \rangle \vert g \rangle\  \leftrightarrow \left[\begin{array}{c} 0 \\ 0 \\ 1 \\ 0 \end{array}\right]
  , \,\, \\
  (12) \vert e \rangle \vert g \rangle\ = \vert g \rangle \vert e \rangle\  \leftrightarrow \left[\begin{array}{c} 0 \\ 1 \\ 0 \\ 0 \end{array}\right]
  , \,\, \\
  (12) \vert e \rangle \vert e \rangle\ = \vert e \rangle \vert e \rangle\  \leftrightarrow \left[\begin{array}{c} 0 \\ 0 \\ 0 \\ 1 \end{array}\right]
  . \,\,
  \end{split}
  \label{eq_2leveloperatoreffect}
  \end{equation}

\item Assuming fermionic molecules, the molecular basis functions are projected to the irrep s[1$^2$], using
\begin{equation}
 \hat{P}_{\rm fermion}=\frac{1}{2}(E - (12)),
 \label{eq_projectors2}
\end{equation}
which gives
 \begin{equation}
 \begin{split}
  \hat{P}_{\rm fermion} \left\lbrace  \left[\begin{array}{c} 1 \\ 0 \\ 0 \\ 0 \end{array}\right] , \left[\begin{array}{c} 0 \\ 1 \\ 0 \\ 0 \end{array}\right] , \left[\begin{array}{c} 0 \\ 0 \\ 1 \\ 0 \end{array}\right] , \left[\begin{array}{c} 0 \\ 0 \\ 0 \\ 1 \end{array}\right]  \right\rbrace =  \\ \left\lbrace  \left[\begin{array}{c} 0 \\ 0 \\ 0 \\ 0 \end{array}\right] , \left[\begin{array}{c} 0 \\ 1/2 \\ -1/2 \\ 0 \end{array}\right] , \left[\begin{array}{c} 0 \\ -1/2 \\ 1/2 \\ 0 \end{array}\right] , \left[\begin{array}{c} 0 \\ 0 \\ 0 \\ 0 \end{array}\right]  \right\rbrace ,
  \end{split}
  \label{eq_operatoreffect}
  \end{equation}
for the (3.a) full molecular state space and
 \begin{equation}
 \begin{split}
  \hat{P}_{\rm fermion} \left\lbrace  \left[\begin{array}{c} 1 \\ 0 \\ 0 \\ 0 \end{array}\right]  \right\rbrace = \left\lbrace  \left[\begin{array}{c} 0 \\ 0 \\ 0 \\ 0 \end{array}\right]   \right\rbrace ,\, \\
  \hat{P}_{\rm fermion} \left\lbrace  \left[\begin{array}{c} 0 \\ 1 \\ 0 \\ 0 \end{array}\right] , \left[\begin{array}{c} 0 \\ 0 \\ 1 \\ 0 \end{array}\right] \right\rbrace = \\ \left\lbrace  \left[\begin{array}{c} 0 \\ 1/2 \\ -1/2 \\ 0 \end{array}\right] , \left[\begin{array}{c} 0 \\ -1/2 \\ 1/2 \\ 0 \end{array}\right] \right\rbrace ,
  \end{split}
  \label{eq_operatoreffect2}
  \end{equation}
for the two sets of vectors when (3.b) only the first excited manifold of the ``cavity radiation + molecules'' system is considered.
\item
In both cases, (3.a) and (3.b), after orthonormalization only one linearly independent state remains:
\begin{equation}
\frac{1}{\sqrt{2}}\left[\begin{array}{c} 0 \\ 1 \\ -1 \\ 0 \end{array}\right] \leftrightarrow \frac{1}{\sqrt{2}} \left( \vert g \rangle \vert e \rangle - \vert e \rangle \vert g \rangle \right),
\end{equation}
therefore, after multiplication by the photon number states, the final Pauli-allowed states read
\begin{equation}
\frac{1}{\sqrt{2}} \left( \vert g \rangle \vert e \rangle \vert N \rangle - \vert e \rangle \vert g \rangle \vert N \rangle \right ),\,\,\, N=0,1,2,...
\end{equation}
for the (4.a) full state space, and
\begin{equation}
\frac{1}{\sqrt{2}} \left( \vert g \rangle \vert e \rangle \vert 0 \rangle - \vert e \rangle \vert g \rangle \vert 0 \rangle \right ).
\end{equation}
for the (4.b) first excited manifold.
\end{enumerate}

\bibliography{main}

\begin{thebibliography}{38}%
\makeatletter
\providecommand \@ifxundefined [1]{%
 \@ifx{#1\undefined}
}%
\providecommand \@ifnum [1]{%
 \ifnum #1\expandafter \@firstoftwo
 \else \expandafter \@secondoftwo
 \fi
}%
\providecommand \@ifx [1]{%
 \ifx #1\expandafter \@firstoftwo
 \else \expandafter \@secondoftwo
 \fi
}%
\providecommand \natexlab [1]{#1}%
\providecommand \enquote  [1]{``#1''}%
\providecommand \bibnamefont  [1]{#1}%
\providecommand \bibfnamefont [1]{#1}%
\providecommand \citenamefont [1]{#1}%
\providecommand \href@noop [0]{\@secondoftwo}%
\providecommand \href [0]{\begingroup \@sanitize@url \@href}%
\providecommand \@href[1]{\@@startlink{#1}\@@href}%
\providecommand \@@href[1]{\endgroup#1\@@endlink}%
\providecommand \@sanitize@url [0]{\catcode `\\12\catcode `\$12\catcode
  `\&12\catcode `\#12\catcode `\^12\catcode `\_12\catcode `\%12\relax}%
\providecommand \@@startlink[1]{}%
\providecommand \@@endlink[0]{}%
\providecommand \url  [0]{\begingroup\@sanitize@url \@url }%
\providecommand \@url [1]{\endgroup\@href {#1}{\urlprefix }}%
\providecommand \urlprefix  [0]{URL }%
\providecommand \Eprint [0]{\href }%
\providecommand \doibase [0]{https://doi.org/}%
\providecommand \selectlanguage [0]{\@gobble}%
\providecommand \bibinfo  [0]{\@secondoftwo}%
\providecommand \bibfield  [0]{\@secondoftwo}%
\providecommand \translation [1]{[#1]}%
\providecommand \BibitemOpen [0]{}%
\providecommand \bibitemStop [0]{}%
\providecommand \bibitemNoStop [0]{.\EOS\space}%
\providecommand \EOS [0]{\spacefactor3000\relax}%
\providecommand \BibitemShut  [1]{\csname bibitem#1\endcsname}%
\let\auto@bib@innerbib\@empty
%</preamble>
\bibitem [{\citenamefont {Fierz}(1939)}]{39Fi}%
  \BibitemOpen
  \bibfield  {author} {\bibinfo {author} {\bibfnamefont {M.}~\bibnamefont
  {Fierz}},\ }\bibfield  {title} {\bibinfo {title} {\"{U}ber die
  relativistische theorie kr\"{a}ftefreier teilchen mit beliebigem spin},\
  }\href@noop {} {\bibfield  {journal} {\bibinfo  {journal} {Helvetica Physica
  Acta}\ }\textbf {\bibinfo {volume} {12}},\ \bibinfo {pages} {3} (\bibinfo
  {year} {1939})}\BibitemShut {NoStop}%
\bibitem [{\citenamefont {Pauli}(1940)}]{40Pa}%
  \BibitemOpen
  \bibfield  {author} {\bibinfo {author} {\bibfnamefont {W.}~\bibnamefont
  {Pauli}},\ }\bibfield  {title} {\bibinfo {title} {The connection between spin
  and statistics},\ }\href@noop {} {\bibfield  {journal} {\bibinfo  {journal}
  {Physical Review}\ }\textbf {\bibinfo {volume} {58}},\ \bibinfo {pages} {716}
  (\bibinfo {year} {1940})}\BibitemShut {NoStop}%
\bibitem [{\citenamefont {Bunker}\ and\ \citenamefont {Jensen}(2006)}]{06BuJe}%
  \BibitemOpen
  \bibfield  {author} {\bibinfo {author} {\bibfnamefont {P.~R.}\ \bibnamefont
  {Bunker}}\ and\ \bibinfo {author} {\bibfnamefont {P.}~\bibnamefont
  {Jensen}},\ }\href@noop {} {\emph {\bibinfo {title} {{Molecular Symmetry and
  Spectroscopy}}}}\ (\bibinfo  {publisher} {NRC Research Press, Ottawa},\
  \bibinfo {year} {2006})\BibitemShut {NoStop}%
\bibitem [{\citenamefont {Kleindienst}\ \emph {et~al.}(1995)\citenamefont
  {Kleindienst}, \citenamefont {L\"{u}chow},\ and\ \citenamefont
  {Barrois}}]{95KlLuBa}%
  \BibitemOpen
  \bibfield  {author} {\bibinfo {author} {\bibfnamefont {H.}~\bibnamefont
  {Kleindienst}}, \bibinfo {author} {\bibfnamefont {A.}~\bibnamefont
  {L\"{u}chow}},\ and\ \bibinfo {author} {\bibfnamefont {R.}~\bibnamefont
  {Barrois}},\ }\bibfield  {title} {\bibinfo {title} {Pauli principle and
  permutation symmetry},\ }\href@noop {} {\bibfield  {journal} {\bibinfo
  {journal} {Journal of Chemical Education}\ }\textbf {\bibinfo {volume}
  {72}},\ \bibinfo {pages} {1019} (\bibinfo {year} {1995})}\BibitemShut
  {NoStop}%
\bibitem [{\citenamefont {Furtenbacher}\ \emph {et~al.}(2013)\citenamefont
  {Furtenbacher}, \citenamefont {Szidarovszky}, \citenamefont {M{\'{a}}tyus},
  \citenamefont {F{\'{a}}bri},\ and\ \citenamefont
  {Cs{\'{a}}sz{\'{a}}r}}]{13FuSzMaFa}%
  \BibitemOpen
  \bibfield  {author} {\bibinfo {author} {\bibfnamefont {T.}~\bibnamefont
  {Furtenbacher}}, \bibinfo {author} {\bibfnamefont {T.}~\bibnamefont
  {Szidarovszky}}, \bibinfo {author} {\bibfnamefont {E.}~\bibnamefont
  {M{\'{a}}tyus}}, \bibinfo {author} {\bibfnamefont {C.}~\bibnamefont
  {F{\'{a}}bri}},\ and\ \bibinfo {author} {\bibfnamefont {A.~G.}\ \bibnamefont
  {Cs{\'{a}}sz{\'{a}}r}},\ }\bibfield  {title} {\bibinfo {title} {Analysis of
  the rotational-vibrational states of the molecular ion {H}$_3^+$},\
  }\href@noop {} {\bibfield  {journal} {\bibinfo  {journal} {Journal of
  Chemical Theory and Computation}\ }\textbf {\bibinfo {volume} {9}},\ \bibinfo
  {pages} {5471} (\bibinfo {year} {2013})}\BibitemShut {NoStop}%
\bibitem [{\citenamefont {Ebbesen}(2016)}]{Ebbesen_review_2016}%
  \BibitemOpen
  \bibfield  {author} {\bibinfo {author} {\bibfnamefont {T.~W.}\ \bibnamefont
  {Ebbesen}},\ }\bibfield  {title} {\bibinfo {title} {{Hybrid Light-Matter
  States in a Molecular and Material Science Perspective}},\ }\href@noop {}
  {\bibfield  {journal} {\bibinfo  {journal} {Accounts of Chemical Research}\
  }\textbf {\bibinfo {volume} {49}},\ \bibinfo {pages} {2403} (\bibinfo {year}
  {2016})}\BibitemShut {NoStop}%
\bibitem [{\citenamefont {Ribeiro}\ \emph {et~al.}(2018)\citenamefont
  {Ribeiro}, \citenamefont {Mart{\'{\i}}nez-Mart{\'{\i}}nez}, \citenamefont
  {Du}, \citenamefont {Campos-Gonzalez-Angulo},\ and\ \citenamefont
  {Yuen-Zhou}}]{Zhou_review_2018}%
  \BibitemOpen
  \bibfield  {author} {\bibinfo {author} {\bibfnamefont {R.~F.}\ \bibnamefont
  {Ribeiro}}, \bibinfo {author} {\bibfnamefont {L.~A.}\ \bibnamefont
  {Mart{\'{\i}}nez-Mart{\'{\i}}nez}}, \bibinfo {author} {\bibfnamefont
  {M.}~\bibnamefont {Du}}, \bibinfo {author} {\bibfnamefont {J.}~\bibnamefont
  {Campos-Gonzalez-Angulo}},\ and\ \bibinfo {author} {\bibfnamefont
  {J.}~\bibnamefont {Yuen-Zhou}},\ }\bibfield  {title} {\bibinfo {title}
  {Polariton chemistry: controlling molecular dynamics with optical cavities},\
  }\href@noop {} {\bibfield  {journal} {\bibinfo  {journal} {Chem. Sci.}\
  }\textbf {\bibinfo {volume} {9}},\ \bibinfo {pages} {6325} (\bibinfo {year}
  {2018})}\BibitemShut {NoStop}%
\bibitem [{\citenamefont {Feist}\ \emph {et~al.}(2018)\citenamefont {Feist},
  \citenamefont {Galego},\ and\ \citenamefont
  {Garcia-Vidal}}]{Feist_GarciaVidal_ACSphotonics_2018}%
  \BibitemOpen
  \bibfield  {author} {\bibinfo {author} {\bibfnamefont {J.}~\bibnamefont
  {Feist}}, \bibinfo {author} {\bibfnamefont {J.}~\bibnamefont {Galego}},\ and\
  \bibinfo {author} {\bibfnamefont {F.~J.}\ \bibnamefont {Garcia-Vidal}},\
  }\bibfield  {title} {\bibinfo {title} {{Polaritonic Chemistry with Organic
  Molecules}},\ }\href@noop {} {\bibfield  {journal} {\bibinfo  {journal} {ACS
  Photonics}\ }\textbf {\bibinfo {volume} {5}},\ \bibinfo {pages} {205}
  (\bibinfo {year} {2018})}\BibitemShut {NoStop}%
\bibitem [{\citenamefont {Chikkaraddy}\ \emph {et~al.}(2016)\citenamefont
  {Chikkaraddy}, \citenamefont {de~Nijs}, \citenamefont {Benz}, \citenamefont
  {Barrow}, \citenamefont {Scherman}, \citenamefont {Rosta}, \citenamefont
  {Demetriadou}, \citenamefont {Fox}, \citenamefont {Hess},\ and\ \citenamefont
  {Baumberg}}]{Edina1}%
  \BibitemOpen
  \bibfield  {author} {\bibinfo {author} {\bibfnamefont {R.}~\bibnamefont
  {Chikkaraddy}}, \bibinfo {author} {\bibfnamefont {B.}~\bibnamefont
  {de~Nijs}}, \bibinfo {author} {\bibfnamefont {F.}~\bibnamefont {Benz}},
  \bibinfo {author} {\bibfnamefont {S.~J.}\ \bibnamefont {Barrow}}, \bibinfo
  {author} {\bibfnamefont {O.~A.}\ \bibnamefont {Scherman}}, \bibinfo {author}
  {\bibfnamefont {E.}~\bibnamefont {Rosta}}, \bibinfo {author} {\bibfnamefont
  {A.}~\bibnamefont {Demetriadou}}, \bibinfo {author} {\bibfnamefont
  {P.}~\bibnamefont {Fox}}, \bibinfo {author} {\bibfnamefont {O.}~\bibnamefont
  {Hess}},\ and\ \bibinfo {author} {\bibfnamefont {J.~J.}\ \bibnamefont
  {Baumberg}},\ }\bibfield  {title} {\bibinfo {title} {Single-molecule strong
  coupling at room temperature in plasmonic nanocavities},\ }\href
  {http://dx.doi.org/10.1038/nature17974} {\bibfield  {journal} {\bibinfo
  {journal} {Nature}\ }\textbf {\bibinfo {volume} {535}},\ \bibinfo {pages}
  {127} (\bibinfo {year} {2016})}\BibitemShut {NoStop}%
\bibitem [{\citenamefont {Herrera}\ and\ \citenamefont
  {Spano}(2016)}]{Herrera_PRL_2016}%
  \BibitemOpen
  \bibfield  {author} {\bibinfo {author} {\bibfnamefont {F.}~\bibnamefont
  {Herrera}}\ and\ \bibinfo {author} {\bibfnamefont {F.~C.}\ \bibnamefont
  {Spano}},\ }\bibfield  {title} {\bibinfo {title} {Cavity-controlled chemistry
  in molecular ensembles},\ }\href@noop {} {\bibfield  {journal} {\bibinfo
  {journal} {Physical Review Letters}\ }\textbf {\bibinfo {volume} {116}},\
  \bibinfo {pages} {238301} (\bibinfo {year} {2016})}\BibitemShut {NoStop}%
\bibitem [{\citenamefont {Hertzog}\ \emph {et~al.}(2019)\citenamefont
  {Hertzog}, \citenamefont {Wang}, \citenamefont {Mony},\ and\ \citenamefont
  {B\"{o}rjesson}}]{Hertzog2019}%
  \BibitemOpen
  \bibfield  {author} {\bibinfo {author} {\bibfnamefont {M.}~\bibnamefont
  {Hertzog}}, \bibinfo {author} {\bibfnamefont {M.}~\bibnamefont {Wang}},
  \bibinfo {author} {\bibfnamefont {J.}~\bibnamefont {Mony}},\ and\ \bibinfo
  {author} {\bibfnamefont {K.}~\bibnamefont {B\"{o}rjesson}},\ }\bibfield
  {title} {\bibinfo {title} {Strong light{\textendash}matter interactions: a
  new direction within chemistry},\ }\href@noop {} {\bibfield  {journal}
  {\bibinfo  {journal} {Chemical Society Reviews}\ }\textbf {\bibinfo {volume}
  {48}},\ \bibinfo {pages} {937} (\bibinfo {year} {2019})}\BibitemShut
  {NoStop}%
\bibitem [{\citenamefont {Herrera}\ and\ \citenamefont
  {Owrutsky}(2020)}]{Herrera_2020}%
  \BibitemOpen
  \bibfield  {author} {\bibinfo {author} {\bibfnamefont {F.}~\bibnamefont
  {Herrera}}\ and\ \bibinfo {author} {\bibfnamefont {J.}~\bibnamefont
  {Owrutsky}},\ }\bibfield  {title} {\bibinfo {title} {Molecular polaritons for
  controlling chemistry with quantum optics},\ }\href@noop {} {\bibfield
  {journal} {\bibinfo  {journal} {The Journal of Chemical Physics}\ }\textbf
  {\bibinfo {volume} {152}},\ \bibinfo {pages} {100902} (\bibinfo {year}
  {2020})}\BibitemShut {NoStop}%
\bibitem [{\citenamefont {Kowalewski}\ and\ \citenamefont
  {Mukamel}(2017)}]{Kowalewski2017}%
  \BibitemOpen
  \bibfield  {author} {\bibinfo {author} {\bibfnamefont {M.}~\bibnamefont
  {Kowalewski}}\ and\ \bibinfo {author} {\bibfnamefont {S.}~\bibnamefont
  {Mukamel}},\ }\bibfield  {title} {\bibinfo {title} {Manipulating molecules
  with quantum light},\ }\href@noop {} {\bibfield  {journal} {\bibinfo
  {journal} {Proceedings of the National Academy of Sciences}\ }\textbf
  {\bibinfo {volume} {114}},\ \bibinfo {pages} {3278} (\bibinfo {year}
  {2017})}\BibitemShut {NoStop}%
\bibitem [{\citenamefont {Herrera}\ and\ \citenamefont
  {Spano}(2017)}]{Herrera2017}%
  \BibitemOpen
  \bibfield  {author} {\bibinfo {author} {\bibfnamefont {F.}~\bibnamefont
  {Herrera}}\ and\ \bibinfo {author} {\bibfnamefont {F.~C.}\ \bibnamefont
  {Spano}},\ }\bibfield  {title} {\bibinfo {title} {Dark vibronic polaritons
  and the spectroscopy of organic microcavities},\ }\href@noop {} {\bibfield
  {journal} {\bibinfo  {journal} {Physical Review Letters}\ }\textbf {\bibinfo
  {volume} {118}} (\bibinfo {year} {2017})}\BibitemShut {NoStop}%
\bibitem [{\citenamefont {Vendrell}(2018)}]{Vendrell2018}%
  \BibitemOpen
  \bibfield  {author} {\bibinfo {author} {\bibfnamefont {O.}~\bibnamefont
  {Vendrell}},\ }\bibfield  {title} {\bibinfo {title} {Collective jahn-teller
  interactions through light-matter coupling in a cavity},\ }\href@noop {}
  {\bibfield  {journal} {\bibinfo  {journal} {Physical Review Letters}\
  }\textbf {\bibinfo {volume} {121}} (\bibinfo {year} {2018})}\BibitemShut
  {NoStop}%
\bibitem [{\citenamefont {Xiang}\ \emph {et~al.}(2018)\citenamefont {Xiang},
  \citenamefont {Ribeiro}, \citenamefont {Dunkelberger}, \citenamefont {Wang},
  \citenamefont {Li}, \citenamefont {Simpkins}, \citenamefont {Owrutsky},
  \citenamefont {Yuen-Zhou},\ and\ \citenamefont {Xiong}}]{Xiang2018}%
  \BibitemOpen
  \bibfield  {author} {\bibinfo {author} {\bibfnamefont {B.}~\bibnamefont
  {Xiang}}, \bibinfo {author} {\bibfnamefont {R.~F.}\ \bibnamefont {Ribeiro}},
  \bibinfo {author} {\bibfnamefont {A.~D.}\ \bibnamefont {Dunkelberger}},
  \bibinfo {author} {\bibfnamefont {J.}~\bibnamefont {Wang}}, \bibinfo {author}
  {\bibfnamefont {Y.}~\bibnamefont {Li}}, \bibinfo {author} {\bibfnamefont
  {B.~S.}\ \bibnamefont {Simpkins}}, \bibinfo {author} {\bibfnamefont {J.~C.}\
  \bibnamefont {Owrutsky}}, \bibinfo {author} {\bibfnamefont {J.}~\bibnamefont
  {Yuen-Zhou}},\ and\ \bibinfo {author} {\bibfnamefont {W.}~\bibnamefont
  {Xiong}},\ }\bibfield  {title} {\bibinfo {title} {Two-dimensional infrared
  spectroscopy of vibrational polaritons},\ }\href@noop {} {\bibfield
  {journal} {\bibinfo  {journal} {Proceedings of the National Academy of
  Sciences}\ }\textbf {\bibinfo {volume} {115}},\ \bibinfo {pages} {4845}
  (\bibinfo {year} {2018})}\BibitemShut {NoStop}%
\bibitem [{\citenamefont {Herrera}\ and\ \citenamefont {Spano}(2018)}]{18HeSp}%
  \BibitemOpen
  \bibfield  {author} {\bibinfo {author} {\bibfnamefont {F.}~\bibnamefont
  {Herrera}}\ and\ \bibinfo {author} {\bibfnamefont {F.~C.}\ \bibnamefont
  {Spano}},\ }\bibfield  {title} {\bibinfo {title} {Theory of nanoscale organic
  cavities: The essential role of vibration-photon dressed states},\
  }\href@noop {} {\bibfield  {journal} {\bibinfo  {journal} {{ACS} Photonics}\
  }\textbf {\bibinfo {volume} {5}},\ \bibinfo {pages} {65} (\bibinfo {year}
  {2018})}\BibitemShut {NoStop}%
\bibitem [{\citenamefont {Csehi}\ \emph {et~al.}(2022)\citenamefont {Csehi},
  \citenamefont {Vendrell}, \citenamefont {Hal{\'{a}}sz},\ and\ \citenamefont
  {Vib{\'{o}}k}}]{22CsVeHaVi}%
  \BibitemOpen
  \bibfield  {author} {\bibinfo {author} {\bibfnamefont {A.}~\bibnamefont
  {Csehi}}, \bibinfo {author} {\bibfnamefont {O.}~\bibnamefont {Vendrell}},
  \bibinfo {author} {\bibfnamefont {G.~J.}\ \bibnamefont {Hal{\'{a}}sz}},\ and\
  \bibinfo {author} {\bibfnamefont {{\'{A}}.}~\bibnamefont {Vib{\'{o}}k}},\
  }\bibfield  {title} {\bibinfo {title} {Competition between collective and
  individual conical intersection dynamics in an optical cavity},\ }\href
  {https://doi.org/10.1088/1367-2630/ac7df7} {\bibfield  {journal} {\bibinfo
  {journal} {New Journal of Physics}\ }\textbf {\bibinfo {volume} {24}},\
  \bibinfo {pages} {073022} (\bibinfo {year} {2022})}\BibitemShut {NoStop}%
\bibitem [{\citenamefont {Sidler}\ \emph {et~al.}(2022)\citenamefont {Sidler},
  \citenamefont {Ruggenthaler}, \citenamefont {Sch\"{a}fer}, \citenamefont
  {Ronca},\ and\ \citenamefont {Rubio}}]{Sidler2022}%
  \BibitemOpen
  \bibfield  {author} {\bibinfo {author} {\bibfnamefont {D.}~\bibnamefont
  {Sidler}}, \bibinfo {author} {\bibfnamefont {M.}~\bibnamefont
  {Ruggenthaler}}, \bibinfo {author} {\bibfnamefont {C.}~\bibnamefont
  {Sch\"{a}fer}}, \bibinfo {author} {\bibfnamefont {E.}~\bibnamefont {Ronca}},\
  and\ \bibinfo {author} {\bibfnamefont {A.}~\bibnamefont {Rubio}},\ }\bibfield
   {title} {\bibinfo {title} {A perspective on ab initio modeling of
  polaritonic chemistry: The role of non-equilibrium effects and quantum
  collectivity},\ }\href@noop {} {\bibfield  {journal} {\bibinfo  {journal}
  {The Journal of Chemical Physics}\ }\textbf {\bibinfo {volume} {156}},\
  \bibinfo {pages} {230901} (\bibinfo {year} {2022})}\BibitemShut {NoStop}%
\bibitem [{\citenamefont {Campos-Gonzalez-Angulo}\ and\ \citenamefont
  {Yuen-Zhou}(2020)}]{CamposGonzalezAngulo_2020}%
  \BibitemOpen
  \bibfield  {author} {\bibinfo {author} {\bibfnamefont {J.~A.}\ \bibnamefont
  {Campos-Gonzalez-Angulo}}\ and\ \bibinfo {author} {\bibfnamefont
  {J.}~\bibnamefont {Yuen-Zhou}},\ }\bibfield  {title} {\bibinfo {title}
  {Polaritonic normal modes in transition state theory},\ }\href@noop {}
  {\bibfield  {journal} {\bibinfo  {journal} {The Journal of Chemical Physics}\
  }\textbf {\bibinfo {volume} {152}},\ \bibinfo {pages} {161101} (\bibinfo
  {year} {2020})}\BibitemShut {NoStop}%
\bibitem [{\citenamefont {Du}\ and\ \citenamefont {Yuen-Zhou}(2022)}]{Du2022}%
  \BibitemOpen
  \bibfield  {author} {\bibinfo {author} {\bibfnamefont {M.}~\bibnamefont
  {Du}}\ and\ \bibinfo {author} {\bibfnamefont {J.}~\bibnamefont {Yuen-Zhou}},\
  }\bibfield  {title} {\bibinfo {title} {Catalysis by dark states in
  vibropolaritonic chemistry},\ }\href@noop {} {\bibfield  {journal} {\bibinfo
  {journal} {Physical Review Letters}\ }\textbf {\bibinfo {volume} {128}},\
  \bibinfo {pages} {096001} (\bibinfo {year} {2022})}\BibitemShut {NoStop}%
\bibitem [{\citenamefont {Tokatly}(2013)}]{Tokatly2013}%
  \BibitemOpen
  \bibfield  {author} {\bibinfo {author} {\bibfnamefont {I.~V.}\ \bibnamefont
  {Tokatly}},\ }\bibfield  {title} {\bibinfo {title} {Time-dependent density
  functional theory for many-electron systems interacting with cavity
  photons},\ }\href@noop {} {\bibfield  {journal} {\bibinfo  {journal}
  {Physical Review Letters}\ }\textbf {\bibinfo {volume} {110}},\ \bibinfo
  {pages} {233001} (\bibinfo {year} {2013})}\BibitemShut {NoStop}%
\bibitem [{\citenamefont {Ruggenthaler}\ \emph {et~al.}(2014)\citenamefont
  {Ruggenthaler}, \citenamefont {Flick}, \citenamefont {Pellegrini},
  \citenamefont {Appel}, \citenamefont {Tokatly},\ and\ \citenamefont
  {Rubio}}]{Ruggenthaler2014}%
  \BibitemOpen
  \bibfield  {author} {\bibinfo {author} {\bibfnamefont {M.}~\bibnamefont
  {Ruggenthaler}}, \bibinfo {author} {\bibfnamefont {J.}~\bibnamefont {Flick}},
  \bibinfo {author} {\bibfnamefont {C.}~\bibnamefont {Pellegrini}}, \bibinfo
  {author} {\bibfnamefont {H.}~\bibnamefont {Appel}}, \bibinfo {author}
  {\bibfnamefont {I.~V.}\ \bibnamefont {Tokatly}},\ and\ \bibinfo {author}
  {\bibfnamefont {A.}~\bibnamefont {Rubio}},\ }\bibfield  {title} {\bibinfo
  {title} {Quantum-electrodynamical density-functional theory: Bridging quantum
  optics and electronic-structure theory},\ }\href@noop {} {\bibfield
  {journal} {\bibinfo  {journal} {Physical Review A}\ }\textbf {\bibinfo
  {volume} {90}},\ \bibinfo {pages} {012508} (\bibinfo {year}
  {2014})}\BibitemShut {NoStop}%
\bibitem [{\citenamefont {Malave}\ \emph {et~al.}(2022)\citenamefont {Malave},
  \citenamefont {Ahrens}, \citenamefont {Pitagora}, \citenamefont {Covington},\
  and\ \citenamefont {Varga}}]{Malave2022}%
  \BibitemOpen
  \bibfield  {author} {\bibinfo {author} {\bibfnamefont {J.}~\bibnamefont
  {Malave}}, \bibinfo {author} {\bibfnamefont {A.}~\bibnamefont {Ahrens}},
  \bibinfo {author} {\bibfnamefont {D.}~\bibnamefont {Pitagora}}, \bibinfo
  {author} {\bibfnamefont {C.}~\bibnamefont {Covington}},\ and\ \bibinfo
  {author} {\bibfnamefont {K.}~\bibnamefont {Varga}},\ }\bibfield  {title}
  {\bibinfo {title} {Real-space, real-time approach to quantum-electrodynamical
  time-dependent density functional theory},\ }\href@noop {} {\bibfield
  {journal} {\bibinfo  {journal} {The Journal of Chemical Physics}\ }\textbf
  {\bibinfo {volume} {157}},\ \bibinfo {pages} {194106} (\bibinfo {year}
  {2022})}\BibitemShut {NoStop}%
\bibitem [{\citenamefont {Sch\"{a}fer}\ \emph {et~al.}(2021)\citenamefont
  {Sch\"{a}fer}, \citenamefont {Buchholz}, \citenamefont {Penz}, \citenamefont
  {Ruggenthaler},\ and\ \citenamefont {Rubio}}]{Schfer2021}%
  \BibitemOpen
  \bibfield  {author} {\bibinfo {author} {\bibfnamefont {C.}~\bibnamefont
  {Sch\"{a}fer}}, \bibinfo {author} {\bibfnamefont {F.}~\bibnamefont
  {Buchholz}}, \bibinfo {author} {\bibfnamefont {M.}~\bibnamefont {Penz}},
  \bibinfo {author} {\bibfnamefont {M.}~\bibnamefont {Ruggenthaler}},\ and\
  \bibinfo {author} {\bibfnamefont {A.}~\bibnamefont {Rubio}},\ }\bibfield
  {title} {\bibinfo {title} {Making ab initio {QED} functional(s):
  Nonperturbative and photon-free effective frameworks for strong light-matter
  coupling},\ }\href@noop {} {\bibfield  {journal} {\bibinfo  {journal}
  {Proceedings of the National Academy of Sciences}\ }\textbf {\bibinfo
  {volume} {118}},\ \bibinfo {pages} {e2110464118} (\bibinfo {year}
  {2021})}\BibitemShut {NoStop}%
\bibitem [{\citenamefont {Haugland}\ \emph {et~al.}(2020)\citenamefont
  {Haugland}, \citenamefont {Ronca}, \citenamefont {Kj{\o}nstad}, \citenamefont
  {Rubio},\ and\ \citenamefont {Koch}}]{Haugland2020}%
  \BibitemOpen
  \bibfield  {author} {\bibinfo {author} {\bibfnamefont {T.~S.}\ \bibnamefont
  {Haugland}}, \bibinfo {author} {\bibfnamefont {E.}~\bibnamefont {Ronca}},
  \bibinfo {author} {\bibfnamefont {E.~F.}\ \bibnamefont {Kj{\o}nstad}},
  \bibinfo {author} {\bibfnamefont {A.}~\bibnamefont {Rubio}},\ and\ \bibinfo
  {author} {\bibfnamefont {H.}~\bibnamefont {Koch}},\ }\bibfield  {title}
  {\bibinfo {title} {Coupled cluster theory for molecular polaritons: Changing
  ground and excited states},\ }\href@noop {} {\bibfield  {journal} {\bibinfo
  {journal} {Physical Review X}\ }\textbf {\bibinfo {volume} {10}},\ \bibinfo
  {pages} {041043} (\bibinfo {year} {2020})}\BibitemShut {NoStop}%
\bibitem [{\citenamefont {Haugland}\ \emph {et~al.}(2021)\citenamefont
  {Haugland}, \citenamefont {Sch\"{a}fer}, \citenamefont {Ronca}, \citenamefont
  {Rubio},\ and\ \citenamefont {Koch}}]{Haugland2021}%
  \BibitemOpen
  \bibfield  {author} {\bibinfo {author} {\bibfnamefont {T.~S.}\ \bibnamefont
  {Haugland}}, \bibinfo {author} {\bibfnamefont {C.}~\bibnamefont
  {Sch\"{a}fer}}, \bibinfo {author} {\bibfnamefont {E.}~\bibnamefont {Ronca}},
  \bibinfo {author} {\bibfnamefont {A.}~\bibnamefont {Rubio}},\ and\ \bibinfo
  {author} {\bibfnamefont {H.}~\bibnamefont {Koch}},\ }\bibfield  {title}
  {\bibinfo {title} {Intermolecular interactions in optical cavities: An ab
  initio {QED} study},\ }\href@noop {} {\bibfield  {journal} {\bibinfo
  {journal} {The Journal of Chemical Physics}\ }\textbf {\bibinfo {volume}
  {154}},\ \bibinfo {pages} {094113} (\bibinfo {year} {2021})}\BibitemShut
  {NoStop}%
\bibitem [{\citenamefont {Cederbaum}(2022)}]{22Ce}%
  \BibitemOpen
  \bibfield  {author} {\bibinfo {author} {\bibfnamefont {L.~S.}\ \bibnamefont
  {Cederbaum}},\ }\bibfield  {title} {\bibinfo {title} {Cooperative molecular
  structure in polaritonic and dark states},\ }\href@noop {} {\bibfield
  {journal} {\bibinfo  {journal} {The Journal of Chemical Physics}\ }\textbf
  {\bibinfo {volume} {156}} (\bibinfo {year} {2022})},\ \bibinfo {note}
  {184102}\BibitemShut {NoStop}%
\bibitem [{\citenamefont {Zeb}\ \emph {et~al.}(2018)\citenamefont {Zeb},
  \citenamefont {Kirton},\ and\ \citenamefont {Keeling}}]{18ZeKiKe}%
  \BibitemOpen
  \bibfield  {author} {\bibinfo {author} {\bibfnamefont {M.~A.}\ \bibnamefont
  {Zeb}}, \bibinfo {author} {\bibfnamefont {P.~G.}\ \bibnamefont {Kirton}},\
  and\ \bibinfo {author} {\bibfnamefont {J.}~\bibnamefont {Keeling}},\
  }\bibfield  {title} {\bibinfo {title} {Exact states and spectra of
  vibrationally dressed polaritons},\ }\href@noop {} {\bibfield  {journal}
  {\bibinfo  {journal} {{ACS} Photonics}\ }\textbf {\bibinfo {volume} {5}},\
  \bibinfo {pages} {249} (\bibinfo {year} {2018})}\BibitemShut {NoStop}%
\bibitem [{\citenamefont {Spano}(2020)}]{20Sp}%
  \BibitemOpen
  \bibfield  {author} {\bibinfo {author} {\bibfnamefont {F.~C.}\ \bibnamefont
  {Spano}},\ }\bibfield  {title} {\bibinfo {title} {Exciton-phonon polaritons
  in organic microcavities: Testing a simple ansatz for treating a large number
  of chromophores},\ }\href@noop {} {\bibfield  {journal} {\bibinfo  {journal}
  {The Journal of Chemical Physics}\ }\textbf {\bibinfo {volume} {152}},\
  \bibinfo {pages} {204113} (\bibinfo {year} {2020})}\BibitemShut {NoStop}%
\bibitem [{\citenamefont {P{\'{e}}rez-S{\'{a}}nchez}\ \emph
  {et~al.}(2023)\citenamefont {P{\'{e}}rez-S{\'{a}}nchez}, \citenamefont
  {Koner}, \citenamefont {Stern},\ and\ \citenamefont
  {Yuen-Zhou}}]{23PeKoStZh}%
  \BibitemOpen
  \bibfield  {author} {\bibinfo {author} {\bibfnamefont {J.~B.}\ \bibnamefont
  {P{\'{e}}rez-S{\'{a}}nchez}}, \bibinfo {author} {\bibfnamefont
  {A.}~\bibnamefont {Koner}}, \bibinfo {author} {\bibfnamefont {N.~P.}\
  \bibnamefont {Stern}},\ and\ \bibinfo {author} {\bibfnamefont
  {J.}~\bibnamefont {Yuen-Zhou}},\ }\bibfield  {title} {\bibinfo {title}
  {Simulating molecular polaritons in the collective regime using few-molecule
  models},\ }\href@noop {} {\bibfield  {journal} {\bibinfo  {journal}
  {Proceedings of the National Academy of Sciences}\ }\textbf {\bibinfo
  {volume} {120}},\ \bibinfo {pages} {e2219223120} (\bibinfo {year}
  {2023})}\BibitemShut {NoStop}%
\bibitem [{\citenamefont {Szidarovszky}\ \emph {et~al.}(2021)\citenamefont
  {Szidarovszky}, \citenamefont {Badank{\'{o}}}, \citenamefont {Hal{\'{a}}sz},\
  and\ \citenamefont {Vib{\'{o}}k}}]{21SzBaHaVi}%
  \BibitemOpen
  \bibfield  {author} {\bibinfo {author} {\bibfnamefont {T.}~\bibnamefont
  {Szidarovszky}}, \bibinfo {author} {\bibfnamefont {P.}~\bibnamefont
  {Badank{\'{o}}}}, \bibinfo {author} {\bibfnamefont {G.~J.}\ \bibnamefont
  {Hal{\'{a}}sz}},\ and\ \bibinfo {author} {\bibfnamefont
  {{\'{A}}.}~\bibnamefont {Vib{\'{o}}k}},\ }\bibfield  {title} {\bibinfo
  {title} {Nonadiabatic phenomena in molecular vibrational polaritons},\
  }\href@noop {} {\bibfield  {journal} {\bibinfo  {journal} {The Journal of
  Chemical Physics}\ }\textbf {\bibinfo {volume} {154}},\ \bibinfo {pages}
  {064305} (\bibinfo {year} {2021})}\BibitemShut {NoStop}%
\bibitem [{\citenamefont {Pauncz}(1995)}]{95Pauncz}%
  \BibitemOpen
  \bibfield  {author} {\bibinfo {author} {\bibfnamefont {R.}~\bibnamefont
  {Pauncz}},\ }\href@noop {} {\emph {\bibinfo {title} {The Symmetric Group in
  Quantum Chemistry}}}\ (\bibinfo  {publisher} {CRC-Press},\ \bibinfo {year}
  {1995})\BibitemShut {NoStop}%
\bibitem [{\citenamefont {Garraway}(2011)}]{Garraway2011}%
  \BibitemOpen
  \bibfield  {author} {\bibinfo {author} {\bibfnamefont {B.~M.}\ \bibnamefont
  {Garraway}},\ }\bibfield  {title} {\bibinfo {title} {The dicke model in
  quantum optics: Dicke model revisited},\ }\href@noop {} {\bibfield  {journal}
  {\bibinfo  {journal} {Philosophical Transactions of the Royal Society A:
  Mathematical, Physical and Engineering Sciences}\ }\textbf {\bibinfo {volume}
  {369}},\ \bibinfo {pages} {1137} (\bibinfo {year} {2011})}\BibitemShut
  {NoStop}%
\bibitem [{\citenamefont {Szidarovszky}(2023)}]{23Sz}%
  \BibitemOpen
  \bibfield  {author} {\bibinfo {author} {\bibfnamefont {T.}~\bibnamefont
  {Szidarovszky}},\ }\bibfield  {title} {\bibinfo {title} {An efficient and
  flexible approach for computing rovibrational polaritons from first
  principles},\ }\href@noop {} {\bibfield  {journal} {\bibinfo  {journal}
  {Journal of Chemical Physics}\ }\textbf {\bibinfo {volume} {159}},\ \bibinfo
  {pages} {014112} (\bibinfo {year} {2023})}\BibitemShut {NoStop}%
\bibitem [{\citenamefont {Furtenbacher}\ \emph {et~al.}(2016)\citenamefont
  {Furtenbacher}, \citenamefont {Szidarovszky}, \citenamefont {Hruby},
  \citenamefont {Kyuberis}, \citenamefont {Zobov}, \citenamefont {Polyansky},
  \citenamefont {Tennyson},\ and\ \citenamefont {Cs\'asz\'ar}}]{16FuSzHrKy}%
  \BibitemOpen
  \bibfield  {author} {\bibinfo {author} {\bibfnamefont {T.}~\bibnamefont
  {Furtenbacher}}, \bibinfo {author} {\bibfnamefont {T.}~\bibnamefont
  {Szidarovszky}}, \bibinfo {author} {\bibfnamefont {J.}~\bibnamefont {Hruby}},
  \bibinfo {author} {\bibfnamefont {A.~A.}\ \bibnamefont {Kyuberis}}, \bibinfo
  {author} {\bibfnamefont {N.~F.}\ \bibnamefont {Zobov}}, \bibinfo {author}
  {\bibfnamefont {O.~L.}\ \bibnamefont {Polyansky}}, \bibinfo {author}
  {\bibfnamefont {J.}~\bibnamefont {Tennyson}},\ and\ \bibinfo {author}
  {\bibfnamefont {A.~G.}\ \bibnamefont {Cs\'asz\'ar}},\ }\bibfield  {title}
  {\bibinfo {title} {{Definitive ideal-gas thermochemical functions of the
  H$_2^{~16}$O molecule}},\ }\href@noop {} {\bibfield  {journal} {\bibinfo
  {journal} {Journal of Physical and Chemical Reference Data}\ }\textbf
  {\bibinfo {volume} {45}},\ \bibinfo {pages} {043104} (\bibinfo {year}
  {2016})}\BibitemShut {NoStop}%
\bibitem [{\citenamefont {Pilar}\ \emph {et~al.}(2020)\citenamefont {Pilar},
  \citenamefont {Bernardis},\ and\ \citenamefont {Rabl}}]{Pilar2020}%
  \BibitemOpen
  \bibfield  {author} {\bibinfo {author} {\bibfnamefont {P.}~\bibnamefont
  {Pilar}}, \bibinfo {author} {\bibfnamefont {D.~D.}\ \bibnamefont
  {Bernardis}},\ and\ \bibinfo {author} {\bibfnamefont {P.}~\bibnamefont
  {Rabl}},\ }\bibfield  {title} {\bibinfo {title} {Thermodynamics of
  ultrastrongly coupled light-matter systems},\ }\href@noop {} {\bibfield
  {journal} {\bibinfo  {journal} {Quantum}\ }\textbf {\bibinfo {volume} {4}},\
  \bibinfo {pages} {335} (\bibinfo {year} {2020})}\BibitemShut {NoStop}%
\bibitem [{\citenamefont {F{\'{a}}bri}(2023)}]{23Fa}%
  \BibitemOpen
  \bibfield  {author} {\bibinfo {author} {\bibfnamefont {C.}~\bibnamefont
  {F{\'{a}}bri}},\ }\bibfield  {title} {\bibinfo {title} {Practical guide to
  the statistical mechanics of molecular polaritons},\ }\bibfield  {journal}
  {\bibinfo  {journal} {Molecular Physics}\ }\href
  {https://doi.org/10.1080/00268976.2023.2272691}
  {10.1080/00268976.2023.2272691} (\bibinfo {year} {2023})\BibitemShut
  {NoStop}%
\end{thebibliography}%

\end{document}